\documentclass{PoS}

\usepackage{epsfig}

\title{Total cross section and particle production in soft and hard processes at the LHC}

\ShortTitle{Total cross section and particle production}

\author{\speaker{Christophe Royon}\thanks{This research is supported by the DOE-Nuclear Physics programme.}\\
        The University of Kansas, Lawrence, USA\\
        E-mail: \email{christophe.royon@ku.edu}}

\abstract{We describe the most recent results at the Large Hadron Collider from the ALICE, ATLAS, CMS, LHCb and TOTEM experiments concerning the elastic, inelastic and total cross section measurements as well as particle production
in soft and hard processes.}

\FullConference{7th Annual Conference on Large Hadron Collider Physics - LHCP2019\\
		20-25 May, 2019\\
		Puebla, Mexico}

\begin{document}

\section{Elastic, inelastic and total cross section measurements}
\begin{figure}
\begin{center}
\epsfig{figure=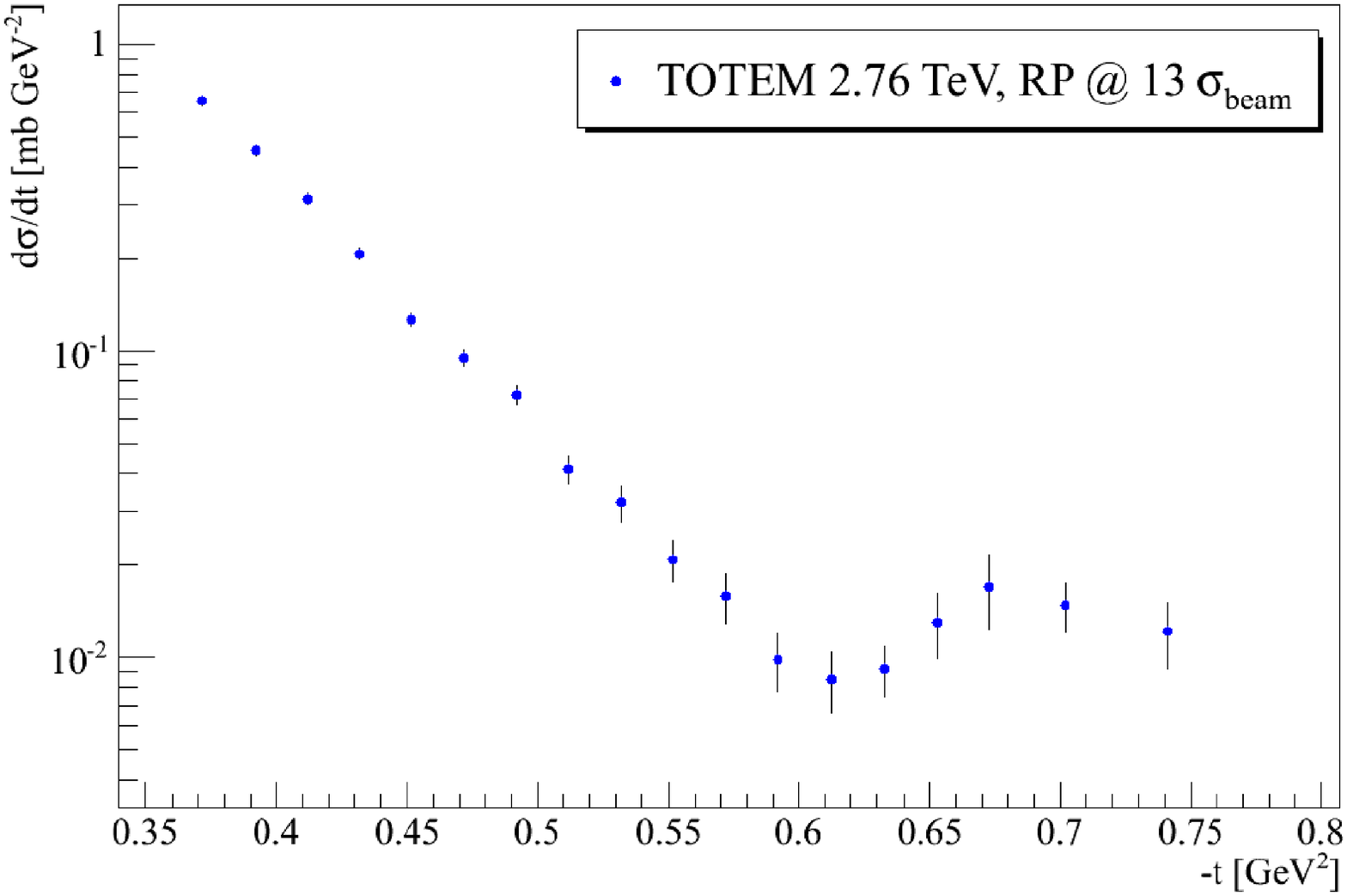,height=0.25\textwidth}
\epsfig{figure=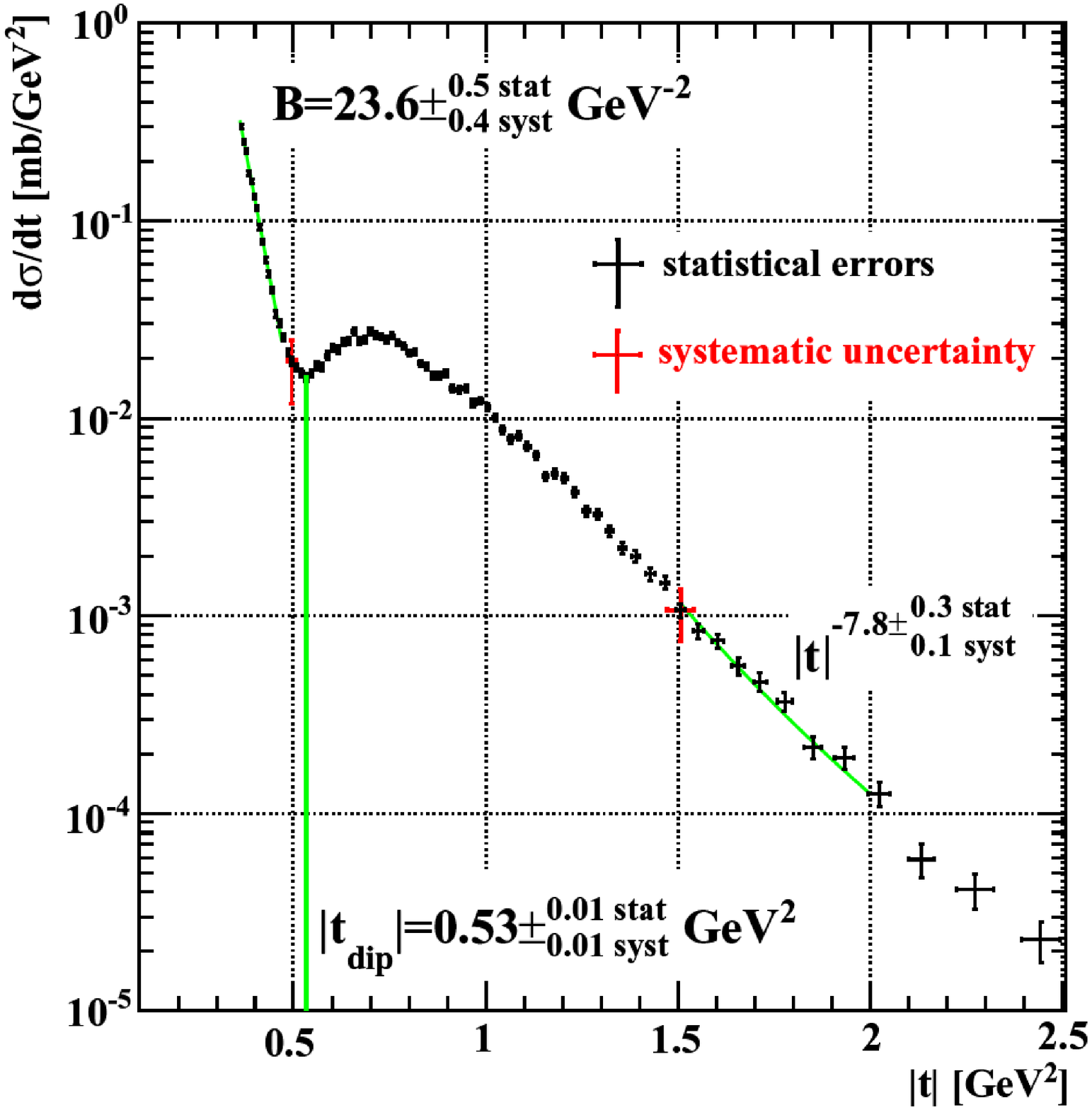,width=0.25\textwidth} 
\epsfig{figure=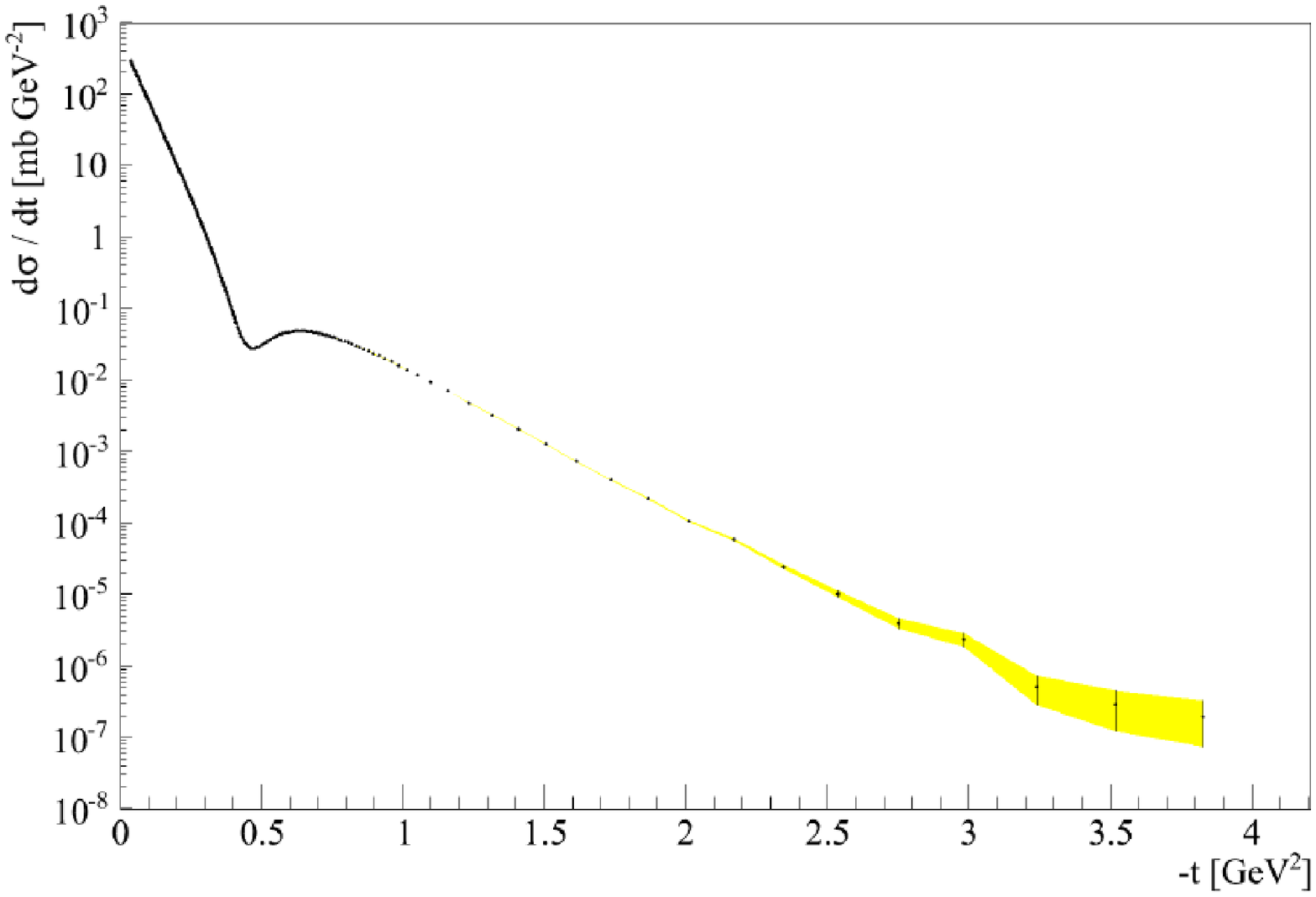,height=0.25\textwidth}
\caption{Elastic $d \sigma/dt$ cross sections measured by the TOTEM collaboration at center-of-mass energies of 2.76, 7 and 13 TeV.}
\label{elastic}
\end{center}
\end{figure}

The ATLAS and TOTEM collaborations measured the elastic $pp \rightarrow pp$ cross sections by detecting both intact
protons in the final state and vetoing on activities in the main CMS or ATLAS detectors. The TOTEM and
ATLAS-ALFA collaborations installed sets of vertical roman pot detectors at about 220-240 meters from the interaction
point in order to detect intact protons in elastic interactions. The elastic event trigger requires the presence of one intact proton on each side of the interaction point on
UP-DOWN or DOWN-UP configurations. In addition to the roman pots detectors, the TOTEM collaboration installed
two inelastic telescopes called $T1$ and $T2$ covering respectively the region $3.1 < |\eta| < 4.7$ and $5.3 < |\eta|
<6.5$ for charged particle $p_T$ above 100 and 40 MeV, respectively. Requesting no activity in $T1$ and $T2$ allows
suppressing 92\% of inelastic background. As an example, the $t$-dependence measurement of the elastic cross section
for center-of-mass energies of 2.76, 7 and 13 TeV is shown in Fig.~\ref{elastic}. We note the presence of a dip and a maximum 
towards $|t| \sim 0.5=0.6$ GeV$^2$ at all center-of-mass energies~\cite{total}. The other noticeable result is that there
is no structure at high $t$ at high center-of-mass energies contrary to what some parametrizations assumed before the
LHC era. The very precise measurements of the data at 8 and 13 TeV
allow the TOTEM collaboration to probe the accuracy of the exponential dependence of elastic data. A simple exponential
fit to $d \sigma/dt$ data at 8 TeV is excluded at 7.2$\sigma$. A fit to $d\sigma/dt$ of the form $A exp (-B(t) |t|)$ with $B(t)$ 
being a linear or a quartic polynomial form leads to a good description of data. The recent 13 TeV elastic cross section
confirm the non-exponential $t$-dependence of elastic data.

\begin{figure}
\begin{center}
\epsfig{figure=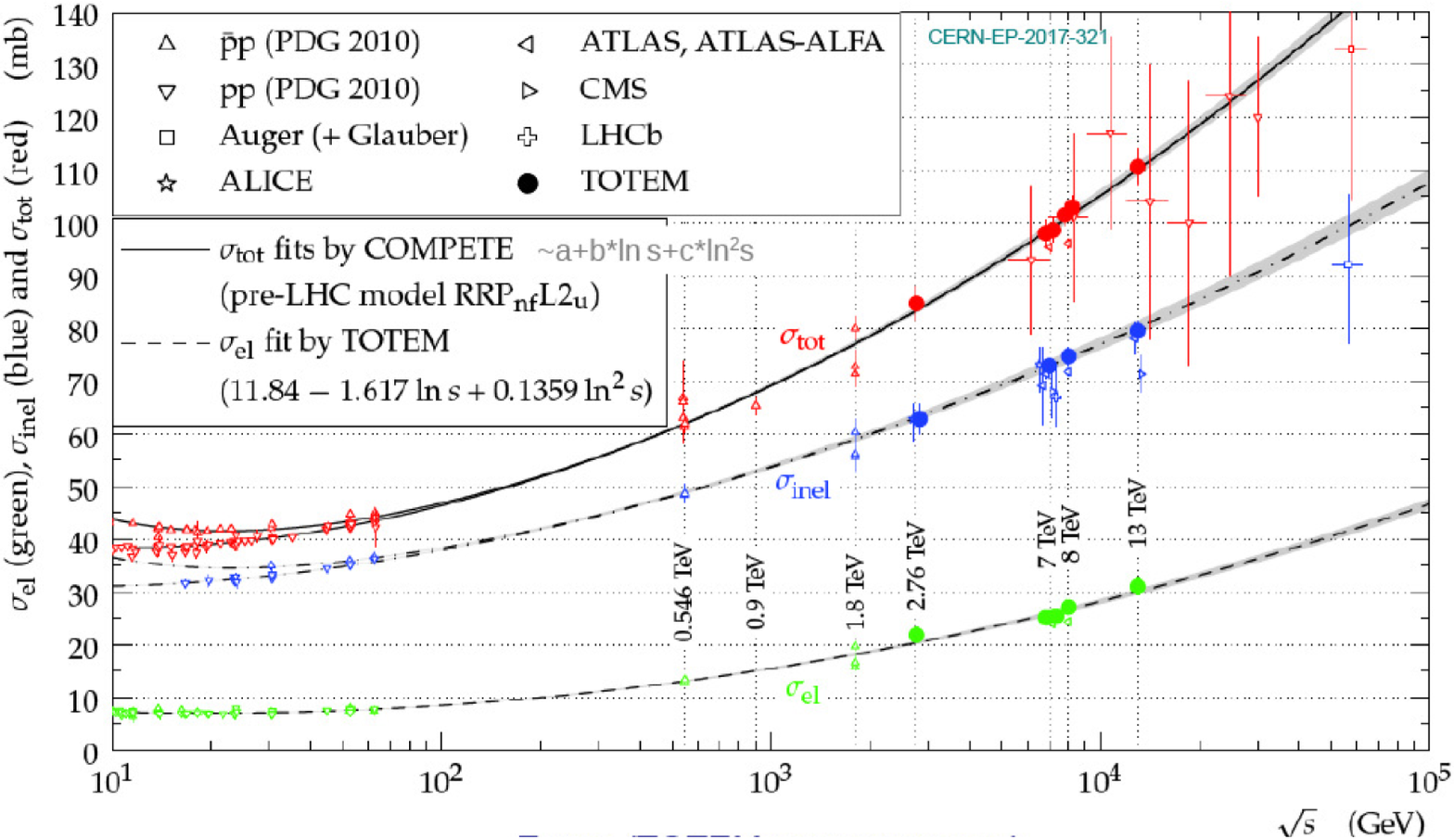,height=0.35\textwidth}
\caption{Elastic, inelastic and total cross sections measured at the LHC by the ALICE, ATLAS, CMS, LHCb and TOTEM
collaborations compared to previous measurements and to fits performed by the TOTEM and COMPETE collaborations}
\label{totalcross}
\end{center}
\end{figure}

\begin{figure}
\begin{center}
\epsfig{figure=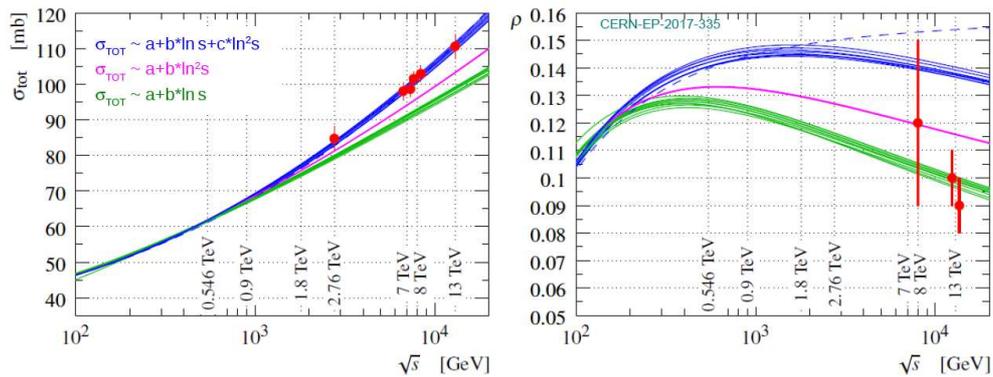,height=0.35\textwidth}
\caption{Measurements of the total cross sections and the $\rho$ parameter as a function of $\sqrt{s}$ compared to 
fits using linear, quartic and a combination of linear and quartic terms.}
\label{rho}
\end{center}
\end{figure}

\begin{figure}
\begin{center}
\epsfig{figure=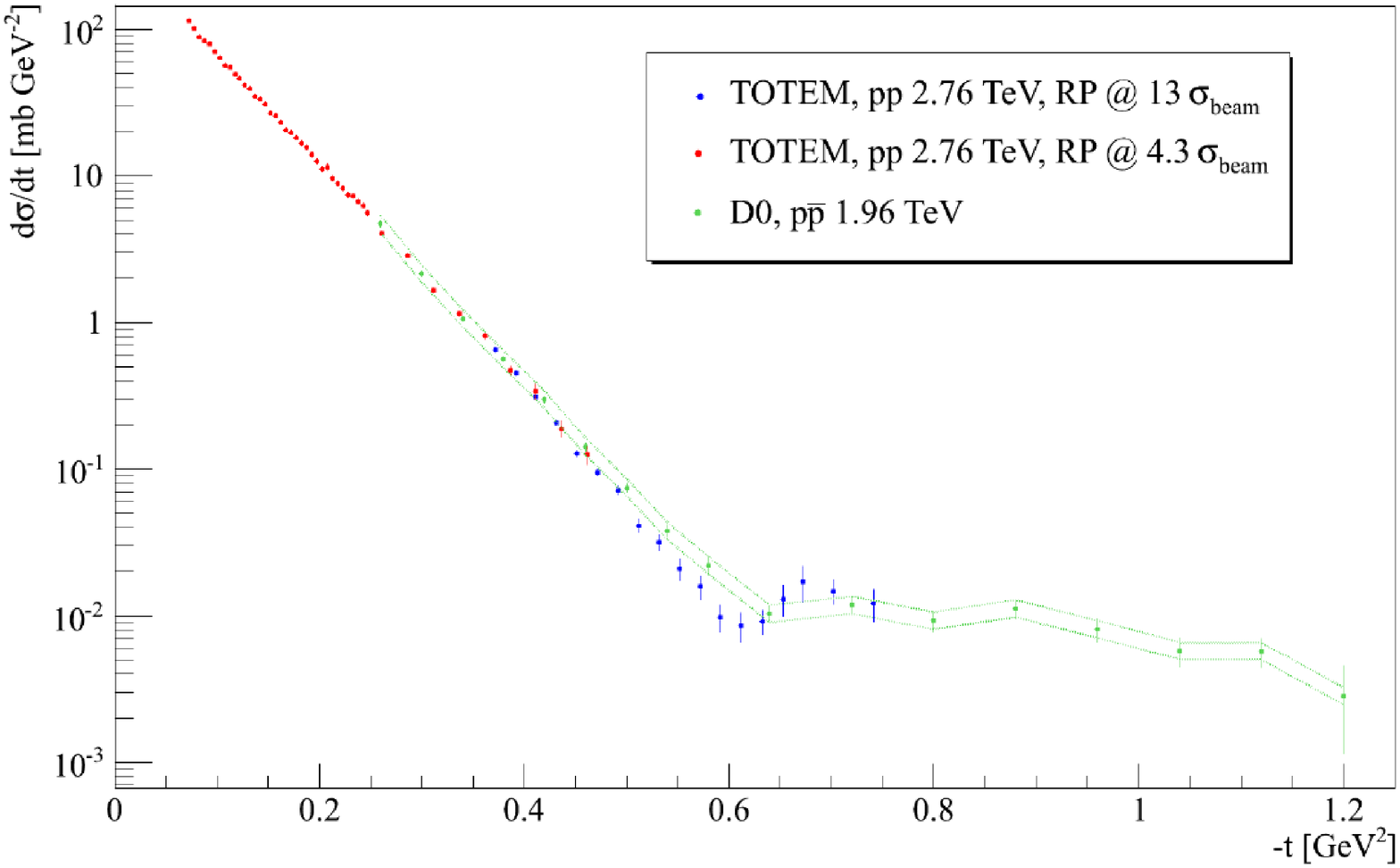,height=0.35\textwidth}
\caption{Comparison between the $pp$ TOTEM cross section measurement at 2.76 Tev and the $p \bar{p}$ measurement from D0 at 1.96 TeV.}
\label{d0}
\end{center}
\end{figure}

The elastic, inelastic and total cross section measurements performed by all LHC experiments are shown in Fig.~\ref{totalcross}. The total cross section at high energies is compatible within uncertainties with previous results from
cosmic ray experiments. A recent measurement of the total cross section from LHCb at 13 TeV is also compatible with
TOTEM results~\cite{lhcbnewcross}. The very precise measurement of the total cross section at 13 TeV down to very 
low $|t|$ allows to measure with high accuracy the $\rho$ parameter, the ratio of the imaginary to the real part of the 
total cross section, $\rho=0.09\pm 0.01$~\cite{totemrho}. The values of the total cross section $\sigma_{tot}$ and the
$\rho$ parameter as a function of $\sqrt{s}$ are shown in Fig.~\ref{rho}. They are compared to a linear, quartic fits
to $\sigma_{tot}$ as a function of $\sqrt{s}$ as well as to a combined linear and quartic fit. $\sigma_{tot}$ Data 
clearly favor the combined fit whereas the $\rho$ measurement at 13 TeV favors the linear fit. The difference between these
two observables can be interpreted as an additional colorless exchange not introduced in these simple fits, namely
the Odderon or 3-gluon exchanges. In order to have better evidence for the existence of the Odderon, it is useful
to compare directly $pp$ and $p \bar{p}$ cross sections. $pp$ and $p \bar{p}$ data from the TOTEM and D0~\cite{d0cross}
are shown in Fig.~\ref{d0}. Even if data were not taken at the same $\sqrt{s}$, it is worth noticing that the $pp$ data
at 2.76 TeV show a dip and maximum  whereas $p \bar{p}$ data do not show such a structure. While some
quantitative studies are still being performed by the D0 and TOTEM collaborations, it is clear that a natural
explanation for the difference between both colliders is due to Odderon exchanges.

In addition, the ATLAS collaboration measured recently single diffractive cross section at 8 TeV. It was noticed that single diffractive
Monte Carlo overestimates the cross section by about a factor 0.64~\cite{atlassd}. Single diffractive data are consistent with an 
exponential $t$ dependence with a slope of 7.60$\pm$0.32 GeV$^{-2}$.

\section{Particle production, energy flow and double parton scattering in $pp$ and heavy ion collisions}
Usually particle production at an hadronic collider is quite complicated. In addition to the hard process involving gluons
and quarks, one needs to take into account initial and final state radiation, underlying events, multi-parton interactions, hadronization effects, 
particle decays... In this section, we will discuss how to understand better these effects using recent measurements
at the LHC.

\subsection{Double parton scattering}
\begin{figure}
\begin{center}
\epsfig{figure=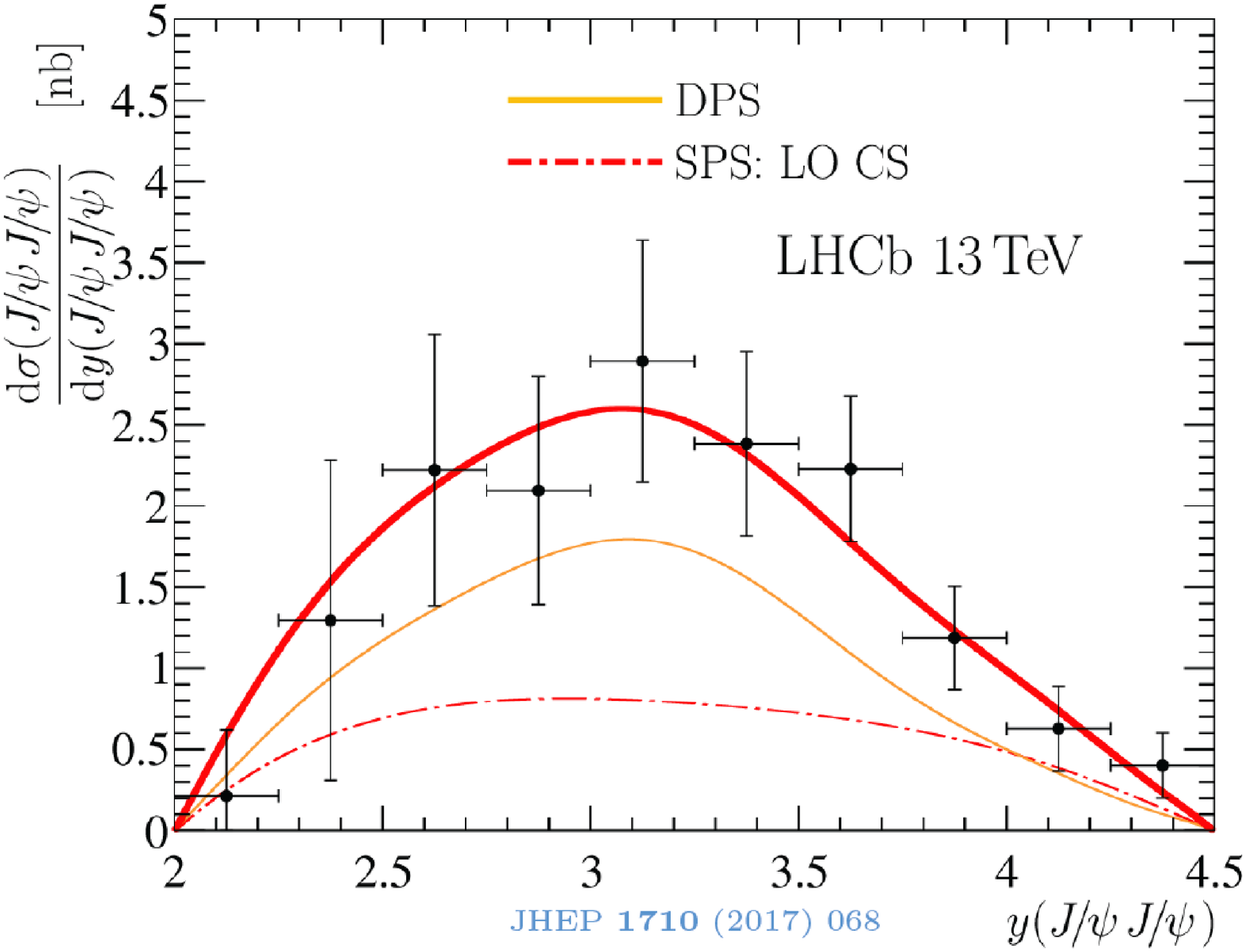,height=0.35\textwidth}
\caption{Strong evidence for DPS in double $J/\Psi$ production by the LHCb collaboration.}
\label{lhcbdps}
\end{center}
\end{figure}

In this section, we will describe recent results at the LHC leading to the observation of double parton scattering (DPS) in 
different channels. DPS is one of the easiest methods to look for multi-patron interactions (MPI) at the LHC. At LHC energies,
MPI  reactions become perturbative. It can be assumed that DPS cross sections factorize 
$\sigma_{AB}^{DPS}=\frac{k}{2}  \frac{\sigma_A \times \sigma_B}{\sigma_{eff}}$ 
where $k$ is a symmetry factor, $\sigma_{A,B}$ are single parton cross sections (SPS), and $\sigma_{eff}$ is the effective overlap area  between hadrons (assumed to be process and $\sqrt{s}$ independent).  In case of the measurement of the 4 charged leptons cross section, one gets
$\sigma_{eff} = \frac{2}{k}  \frac{1}{f_{DPS}} \frac{\sigma^{2l}_A \sigma^{2l}_B}{\sigma^{4l}}$.
The ATLAS collaboration looked for DPS in the 4 lepton channel~\cite{atlasdps} and obtained a DPS fraction consistent
with 0, that leads to an effective overlap area $\sigma_{eff} \ge 1.0$ mb at 95\% C.L., consistent with DPS hypotheses. The LHCb
collaboration looked for DPS in double $J/\Psi$ production at 13 TeV and showed strong evidence for DPS~\cite{lhcbjpsi} as shown in 
Fig.~\ref{lhcbdps}. The CMS collaboration also found clear evidence of DPS using same sign $W$ production at 13 TeV for the first time~\cite{cmsww}.  Combining the
same sign di-muon and electron/muon channels, the CMS collaboration estimated the DPS cross section to be $1.41 \pm 0.28 (stat) \pm 0.28 (syst)$ pb extrapolated to inclusive signal  (i.e. including opposite sign)
that is consistent with DPS processes.

\subsection{Measurement of energy flow in the very forward direction}

\begin{figure}
\begin{center}
\epsfig{figure=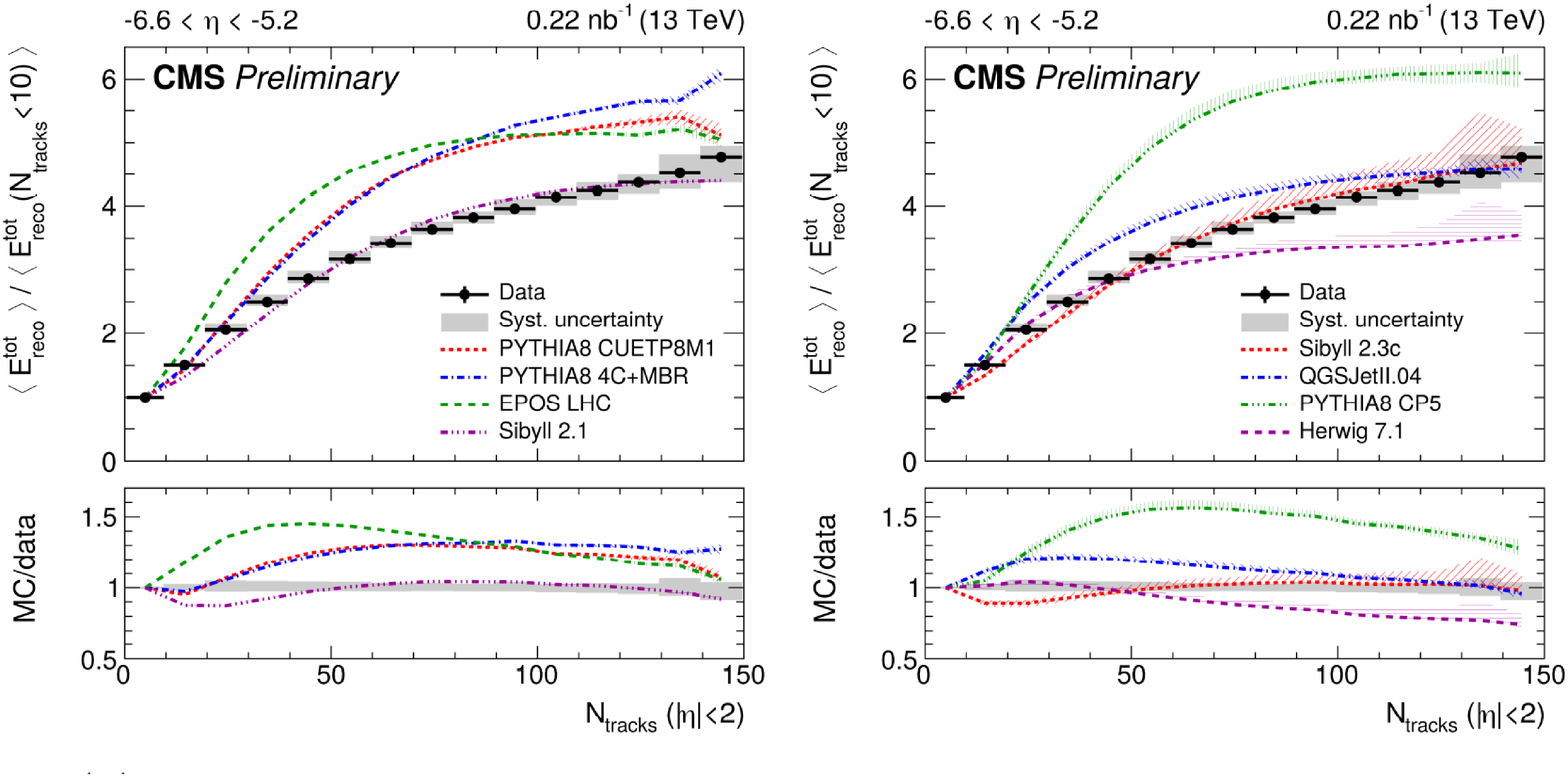,height=0.45\textwidth}
\caption{Measurement of the very forward energy flow using the CASTOR calorimeter of the CMS collaboration.}
\label{cmsflow}
\end{center}
\end{figure}

The CMS collaboration benefitted from the very forward CASTOR calorimeter covering $-6.6 \le \eta \le -5.2$ to measure the energy flow
in the very forward direction~\cite{cmscastor}.  The comparison between the measured energy flow and the Monte Carlo
expectations is given in Fig.~\ref{cmsflow}, where $E_{reco}^{tot}$ 
is obtained by summing over all towers $E_{reco}^{tot} = \Sigma_{i=towers} E_i$. The average energy increases with $N_{tracks}$ and
only  SIBYLL~\cite{sibyll}  and HERWIG~\cite{herwig} describe the relative increase, while EPOS~\cite{epos} and PYTHIA8~\cite{pythia8} fail.

\subsection{Measurement of underlying events and particle production in $pp$ interactions}
The ATLAS collaboration measured underlying events (UE) in $Z+$jet events at 13 TeV. 
They consider four regions in
azimuthal angles, namely the region of the $Z$ boson, of the jet, and the two regions perpendicular. They measure the charged
particle spectrum in the perpendicular region with less activities in order to be sensitive to UE.  All generators produce a higher
fraction of particles at lowest $p_T$ and HERWIG leads to the best results at high $p_T$.

\subsection{Measurement of charged hadron production in $Z$-tagged jets}

\begin{figure}
\begin{center}
\epsfig{figure=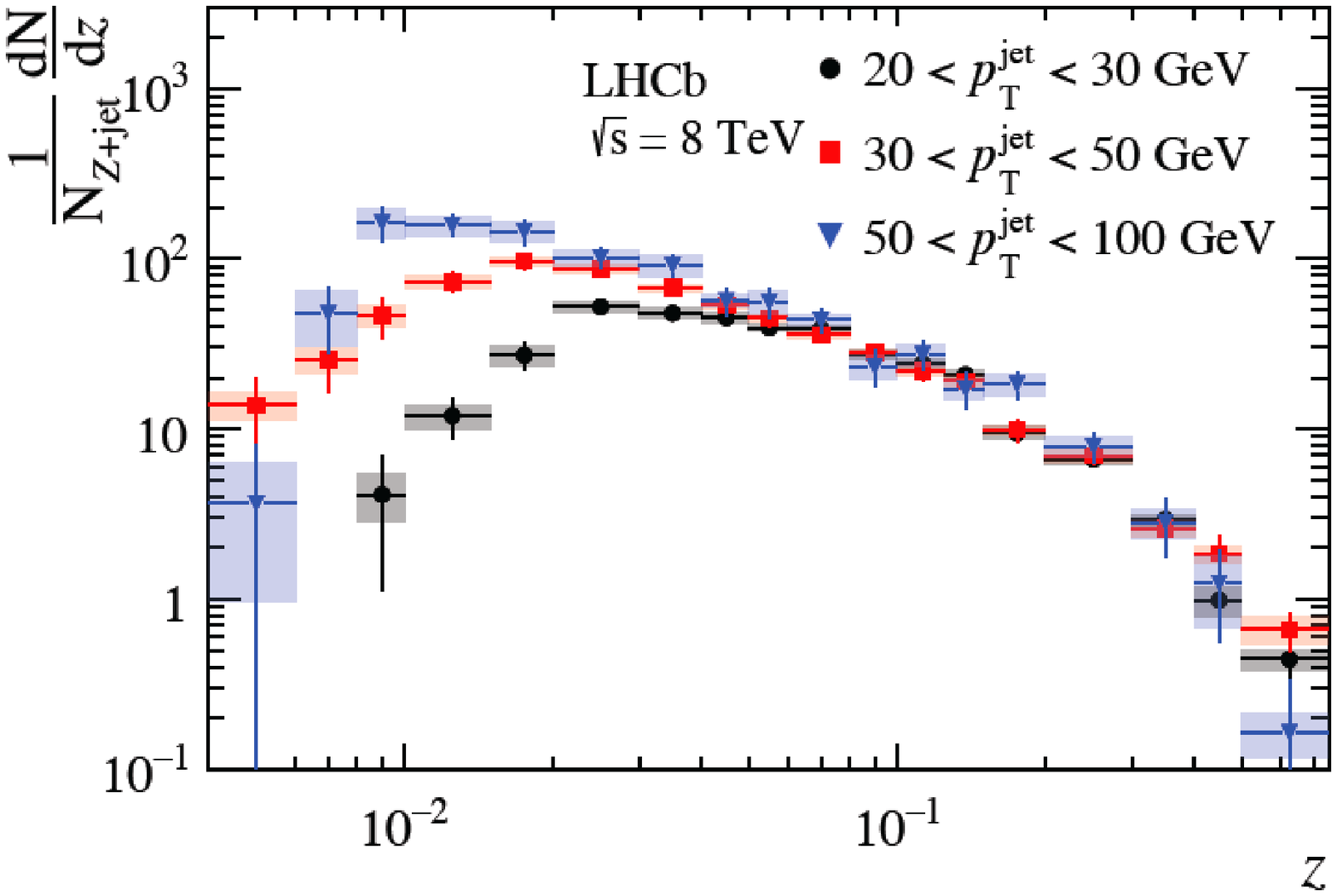,height=0.33\textwidth}
\epsfig{figure=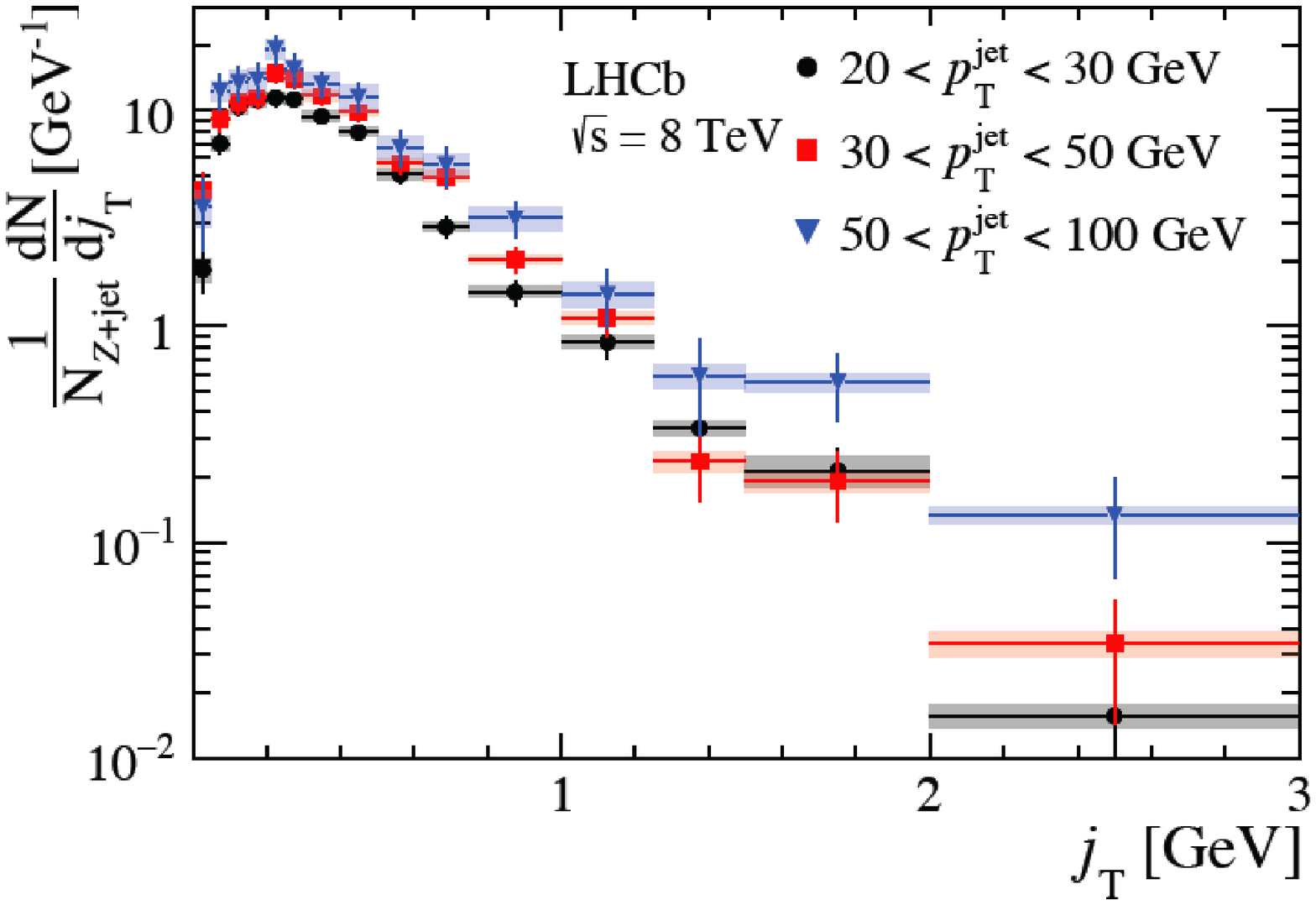,height=0.33\textwidth}
\caption{Charged hadron production in $Z$-tagged jets at 8 TeV.}
\label{lhcbzjet}
\end{center}
\end{figure}

The LHCb collaboration measured recently the charged hadron production in $Z$-tagged jets at a center-of-mass energy of 
8 TeV~\cite{lhcbztag}.  This is the first measurement of jet hadronization at forward rapidities, the jet being produced with the $Z$ 
boson. The measurement probes predominantly light-quark jets and is complementary to previous measurements performed
by other experiments. The results are shown in Fig.~\ref{lhcbzjet} as a function of $z$, the longitudinal momentum fraction of the hadron with respect to the jet) and $j_T$, the transverse momentum of charged hadrons with respect to jet axis. In general. PYTHIA underestimates the 
number of high momentum hadrons within those jets. 

\subsection{Particle production in $pp$ and heavy ion collisions}
\begin{figure}
\begin{center}
\epsfig{figure=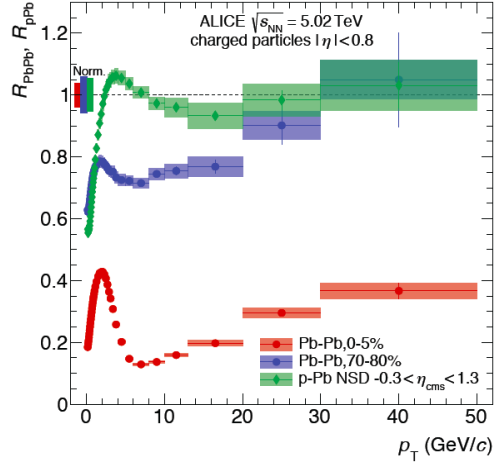,height=0.45\textwidth}
\caption{$p_T$ spectrum of charged particle prodution in $pPb$ and $Pb Pb$ collisions for different centrality values.}
\label{alicept}
\end{center}
\end{figure}

The ALICE collaboration measured the $p_T$ spectrum of particle production in $Pb Pb$ and $Xe Xe$ collisions in 9 different 
centrality bins~\cite{alice}. Depending on the heavy ions and on the centrality, the particle multiplicity can
vary between 5 and 2000 per unit of rapidity. The Alice collaboration performed a full set of measurement of light flavor hadrons 
($\pi$, $K$, $p$, $K_S$, $\Lambda$, $\Xi$, $\Phi$, $\Omega$) as a function of charged particle multiplicity and remarkable similarities 
between hadrons have been observed. The suppression factors in $Pb Pb$ and $p Pb$ collisions are shown in
Fig.~\ref{alicept}. We observe a strong suppression in $Pb Pb$ interactions in central collisions whereas suppression is low in peripheral
collisions or in $p Pb$ interactions, as expected for final state effects. 

\section{Exclusive production of di-leptons and di-photons and prospects}

\subsection{Observation of exclusive di-muon production}
\begin{figure}
\begin{center}
\epsfig{figure=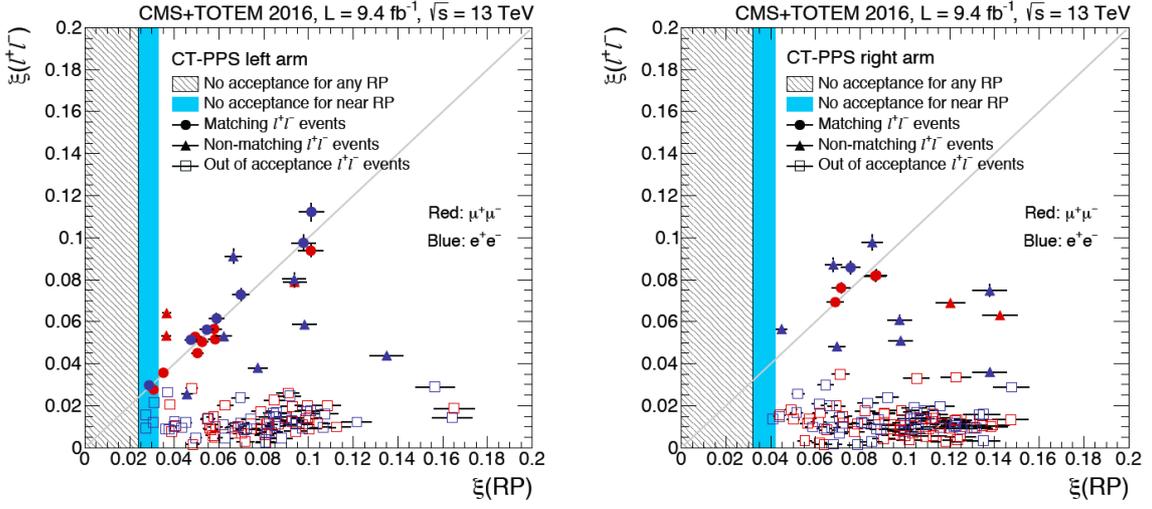,height=0.45\textwidth}
\caption{Correlation between the values of $\xi$ as measured using the di-muon system or the intact protons on both sides of CMS.
The semi-exclusive candidates are shown along the diagonal.}
\label{ctpps}
\end{center}
\end{figure}

The ATLAS collaboration measured the di-muon exclusive diffractive production in $pp$ interactions using special runs at low luminosity~\cite{atlasexclmu}
and the rapidity gap technique (without tagging the protons in the final state).
Results are compatible with SM expectations taking into account proton dissociative effects of the order of 20\%.
The CMS and TOTEM collaborations used the Proton Precision Spectrometer (PPS) to measure the exclusive di-lepton
(di-muon and di-electron) production. Only one proton tag was requested (the mass acceptance of requesting two tagged protons in PPS 
starts at about 450 GeV, and very few events are expected above this mass since the exclusive di-lepton cross section is very low above 450 GeV).
20 semi-exclusive di-lepton candidate events were observed (12 $\mu \mu$ and 8 $ee$) for an estimated background of
$1.49 \pm 0.07$ $(stat)$ $\pm 0.53 (syst)$ in $\mu \mu$  and of $2.36 \pm 0.09$ $(stat)$ $\pm 0.47 (syst)$ in $ee$, which
leads to a discovery of this exclusive process at high di-lepton masses with a significance of about 5.1 $\sigma$~\cite{pps}.
In Fig.~\ref{ctpps}, we show the correlation of the proton fraction momentum loss  $\xi$ measured using the di-lepton system or the
intact proton. Exclusive di-lepton events appear on the diagonal since the value of $\xi$ should be the same within experimental
uncertainties.

\subsection{Observation of exclusive di-photon production and prospects looking for quartic anomalous couplings and axion-like particles}
The ATLAS collaboration measured for the first time the exclusive production of di-photons in $Pb Pb$ collisions at a center-of-mass energy
of 5.02 TeV in 2015 and 2018~\cite{atlasgamma}. Using a luminosity of 1.7 nb$^{-1}$ in 2018, 
59 events were observed for a background of 12$\pm$3.

\begin{figure}
\begin{center}
\epsfig{figure=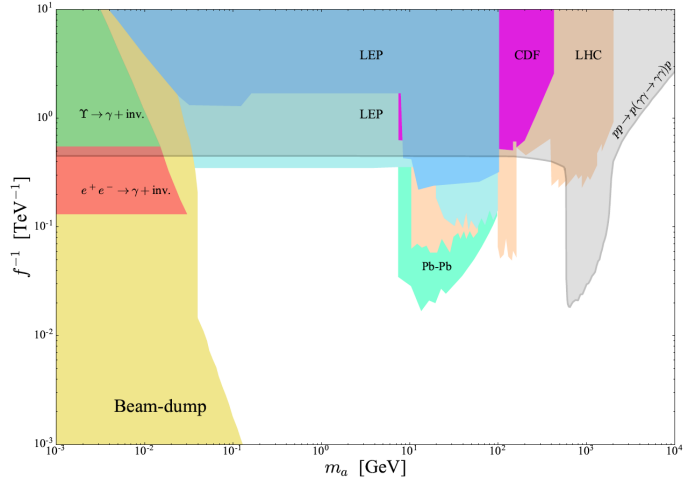,height=0.45\textwidth}
\caption{Sensitivity to ALPs in the coupling versus mass plane using exclusive di-photon events at the LHC.}
\label{alp}
\end{center}
\end{figure}

Using the full luminosity delivered at the LHC and looking for the exclusive production of di-photons at high mass by measuring both the photons in CMS and ATLAS and tagging the 
intact protons in PPS and AFP in the final state allow increasing the sensitivity to quartic $\gamma \gamma \gamma \gamma$ anomalous couplings 
by about two orders of magnitude at the LHC with 300 fb$^{-1}$ compared to other methods~\cite{ourgamma}.  After requesting two high $p_T$ 
and high mass $\gamma$s, the only background that needs to be considered are pile-up events where two additional protons  originating from secondary interactions are in addition 
to the two photons. Requesting a matching between the two proton and di-photon mass and rapidity allows suppressing completely
the background~\cite{ourgamma}. The same method can be applied to look for axion-like particles at high mass~\cite{ouralps}.
The sensitivity plot in the coupling versus mass plane is shown in Fig.~\ref{alp} and we see the increased sensitivities by two orders
of magnitude at high axion-like particle masses using 300 fb$^{-1}$ of $pp$ interactions t 13 TeV compared to present sensitivities.
The same method can be applied to search for quartic $\gamma \gamma WW$ , $\gamma \gamma \gamma Z$, $\gamma \gamma ZZ$
anomalous couplings gaining between two and three orders of magnitude on sensitivity with respect to more standard methods
at the LHC (for instance looking for the decay of the $Z$ boson into three $\gamma$s to look for anomalous $\gamma \gamma \gamma Z$
couplings)~\cite{ourww}.

\section{Conclusion}
In this short review, we started by discussing the most recent results concerning the elastic, inelastic, and total cross section that leads to the 
discrepancy between the total cross section and $\rho$ measurements that can be interpreted as some evidence of three gluon exchange or the odderon. The comparison between the D0 $p \bar{p}$ and TOTEM $pp$ data will be crucial to 
strengthen  the evidence for the odderon. The study of multi-parton interactions and its evidence was performed by many
collaborations at the LHC in double $J/\Psi$ and same sign $W$ analyses. Soft particle production and energy flow were measured in $pp$ and 
heavy ion collisions at the LHC and allow further tuning of Monte Carlo generators. We finished the report by mentioning the observation
of exclusive di-lepton at high mass and the prospects concerning the search for new physics via the possible existence of quartic anomalous couplings
and axion-like particles with unprecedented precisions at the LHC by tagging the intact protons in the final state.

\end{document}